\newcommand{\FF}{{\mathcal F}}

\newcommand{\bea}{\begin{eqnarray}}
\newcommand{\eea}{\end{eqnarray}}

\newcommand{\nn}{\nonumber \\}
\newcommand{\nnn}{\nonumber}
\newcommand{\be}{\begin{eqnarray}}
\newcommand{\ba}{\begin{array}}
\newcommand{\ea}{\end{array}}
\newcommand{\ee}{\end{eqnarray}}

\newcommand{\Tr}{{\rm Tr}}

\documentclass{ws-procs9x6}

\begin{document}

\title{Transversity of quarks and nucleons in SIDIS and Drell Yan
}

\author{Leonard P. Gamberg}

\address{Physics Department,
Penn State-Berks, 
Reading, PA 19610, USA \\
E-mail: lpg10@psu.edu}
\maketitle

\abstracts{
We consider the leading and sub-leading 
twist $T$-odd and even contributions to the 
$\cos 2\phi$ azimuthal asymmetry in unpolarized
dilepton production in Drell-Yan Scattering and semi-inclusive
deep inelastic scattering of pions.
}

\begin{figure}[t]
\includegraphics[width=4.0cm]{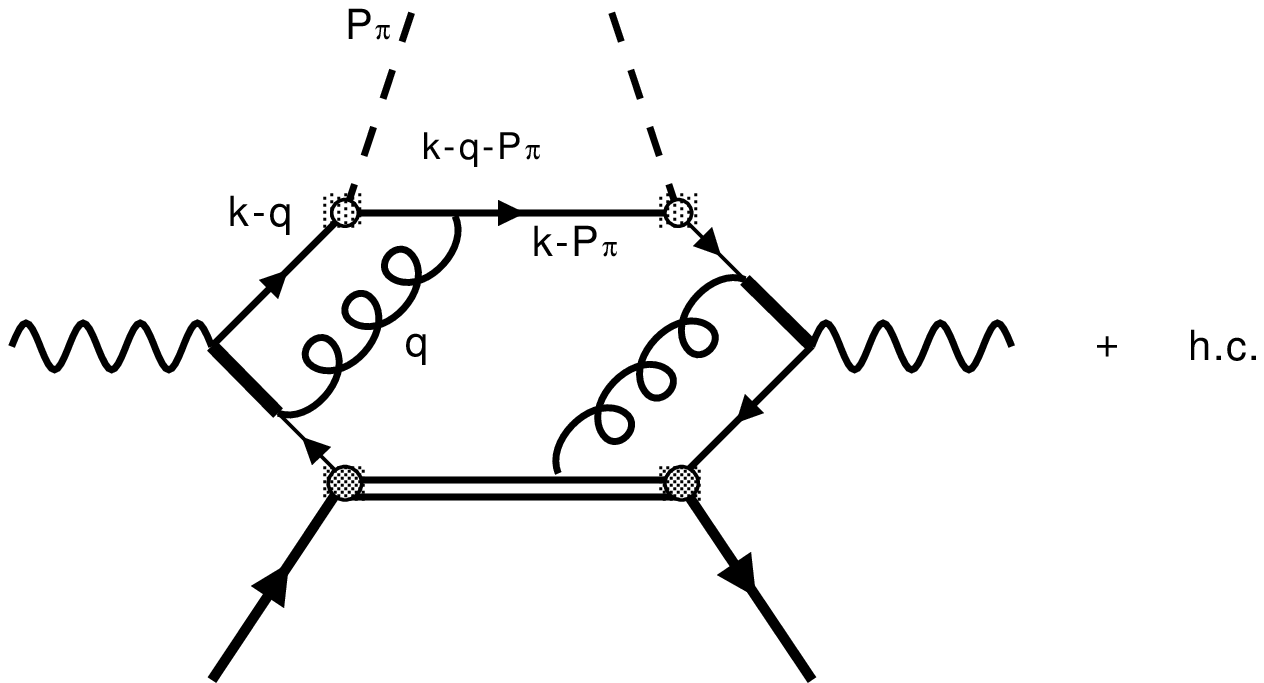}
\includegraphics[width=3.0cm]{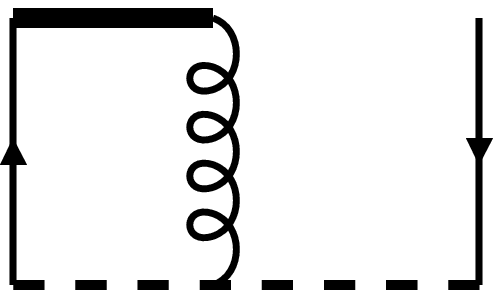}
\includegraphics[width=4.0cm]{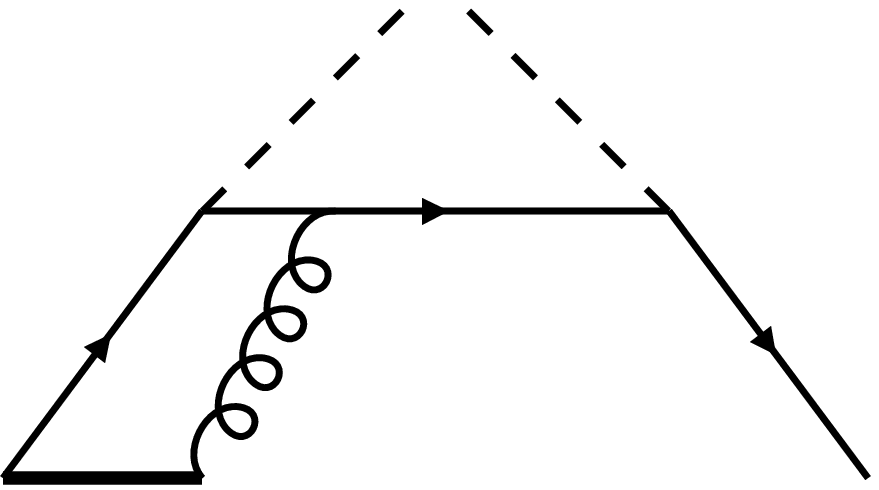}
\caption{\scriptsize   Feynman diagram representing 
initial and final state interactions giving rise
to $T$-odd contribution to quark distribution and fragmentation functions.}
\label{analyze}
\end{figure}

One of the persistent challenges confronting the QCD 
parton model
is the explanation of the significant azimuthal and single spin asymmetries
that emerge  in inclusive and semi-inclusive 
processes~\cite{heller,E615,E704,star,zeus,hermes}.
Going beyond the collinear approximation in PQCD recent progress has been
achieved in
characterizing  these asymmetries in terms 
of absorptive scattering~\cite{bhs,jiyuan,gg}.  Such 
asymmetries involve time-reversal-odd ($T$-odd) 
transverse momentum dependent (TMD) distribution and fragmentation
functions~\cite{sivers,colnpb}.  They are indicative of 
correlations between 
transverse momentum of quarks and or hadrons, and the
transverse spin of the reacting particles~\cite{ansel}.
In SIDIS 
for unpolarized target
$\bm{\mathsf s}_T\cdot\left({\bm P}\times {\bm k}_\perp\right)$,
depicts a correlation of transverse 
quark polarization with the proton's momentum and the
intrinsic quark momentum in an unpolarized 
nucleon while  $\bm{\mathsf s }_T\cdot(\bm{p}\times \bm{P}_{h\perp})$, 
corresponds to that of a fragmenting quark's polarization
with quark and transverse pion momentum, $\bm{P}_{h\perp}$.
These correlations 
enter the {\em unpolarized} 
cross-section  convoluted with $h_1^\perp$~\cite{boermul} and the Collins
fragmentation function $H_1^\perp$~\cite{colnpb}.
 This $\cos 2\phi$  asymmetry is not suppressed by $1/Q$ where $Q$ represents
the hard QCD scale. Model estimates of absorptive scattering 
have led to color gauge invariant~\cite{jiyuan,colplb,boerpij} 
definitions of unsubtracted~\cite{colact,jima} transverse 
momentum dependent distribution and fragmentation
functions.  $h_1^\perp$ is projected from the correlation function for 
TMD distribution functions $\Phi(k,P)$, 
$
\frac{1}{2}\int dk^-\Tr\left(\sigma^{\perp +}\gamma_5\Phi(k,P)\right)=\dots 
\frac{\varepsilon_{
\scriptscriptstyle 
+-\perp j}k_{\scriptscriptstyle \perp j}}{M}h_1^\perp(x,k_\perp)
\dots \quad  .
$
Similarly, the Collins function, $H_{1}^\perp(z,\bm{p}_\perp)$ is projected 
from the fragmentation matrix $\Delta(p,P_h)$.   
Using a parton model within the quark diquark spectator framework to 
model the quark-hadron 
interactions~\cite{gg,ggoprh}
{\small
$$h_1^\perp(x,k_\perp)\hspace{-.10cm} =\hspace{-.10cm}
\frac{{\mathcal N}(m+xM)(1-x)e^{-2b\left(k^2_\perp- \Lambda(0)\right)}}
{\Lambda(k^2_\perp)k_\perp^2}
\hspace{-.15cm}\left[\Gamma(0,2b\Lambda(0))\hspace{-.10cm}-\hspace{-.10cm}
\Gamma(0,2b\Lambda(k^2_\perp))\right]\, ,
$$}
and the gauge link contribution to the Collins function are 
given by~\cite{ggoprf}
{\small
\bea
H_1^\perp(z,k_\perp)=\hspace{-.15cm} 
\frac{{\mathcal N}^\prime\ M_h \mu}{\Lambda(k^2_\perp)k_\perp^2}\frac{(1-z)
e^{-2c\left(k^2_\perp- \Lambda(0)\right)}}{4z^3}
\hspace{-.15cm}\left[\Gamma(0,2c\Lambda(0))\hspace{-.10cm}-\hspace{-.10cm}
\Gamma(0,2c\Lambda(k^2_\perp))\right] .
\nnn\eea
}
Their  contribution to the double $T$-odd azimuthal
$\cos 2\phi$ asymmetry,
in terms of 
initial and final (ISI/FSI) 
state interactions of active or ``struck'', and  fragmenting 
quark~\cite{ggoprh,ggoprf} 
are depicted in Fig.~\ref{analyze}.
Also, it was recognized by Boer~\cite{boerich} that the  $\cos 2\phi$ 
azimuthal asymmetry in 
dilepton production in Drell Yan scattering
 has a $T$-odd 
contribution at leading  twist~\cite{bbh}.
The latter process is interesting in light of  
proposed experiments at Darmstadt  GSI~\cite{pax},
where an anti-proton beam is an ideal for studying 
transversity property of quarks due to the dominance of valence quark 
effects~\cite{hix}.  We explore the role that $T$-odd 
distribution and fragmentation functions play in unpolarized azimuthal 
asymmetries in  SIDIS~\cite{ggoprh,ggoprf,hix}.

\begin{figure}[h]
\begin{center}
\includegraphics[width=5.0cm]{nu_x_50.eps}
\hskip 0.25cm
\includegraphics[width=4.0cm]{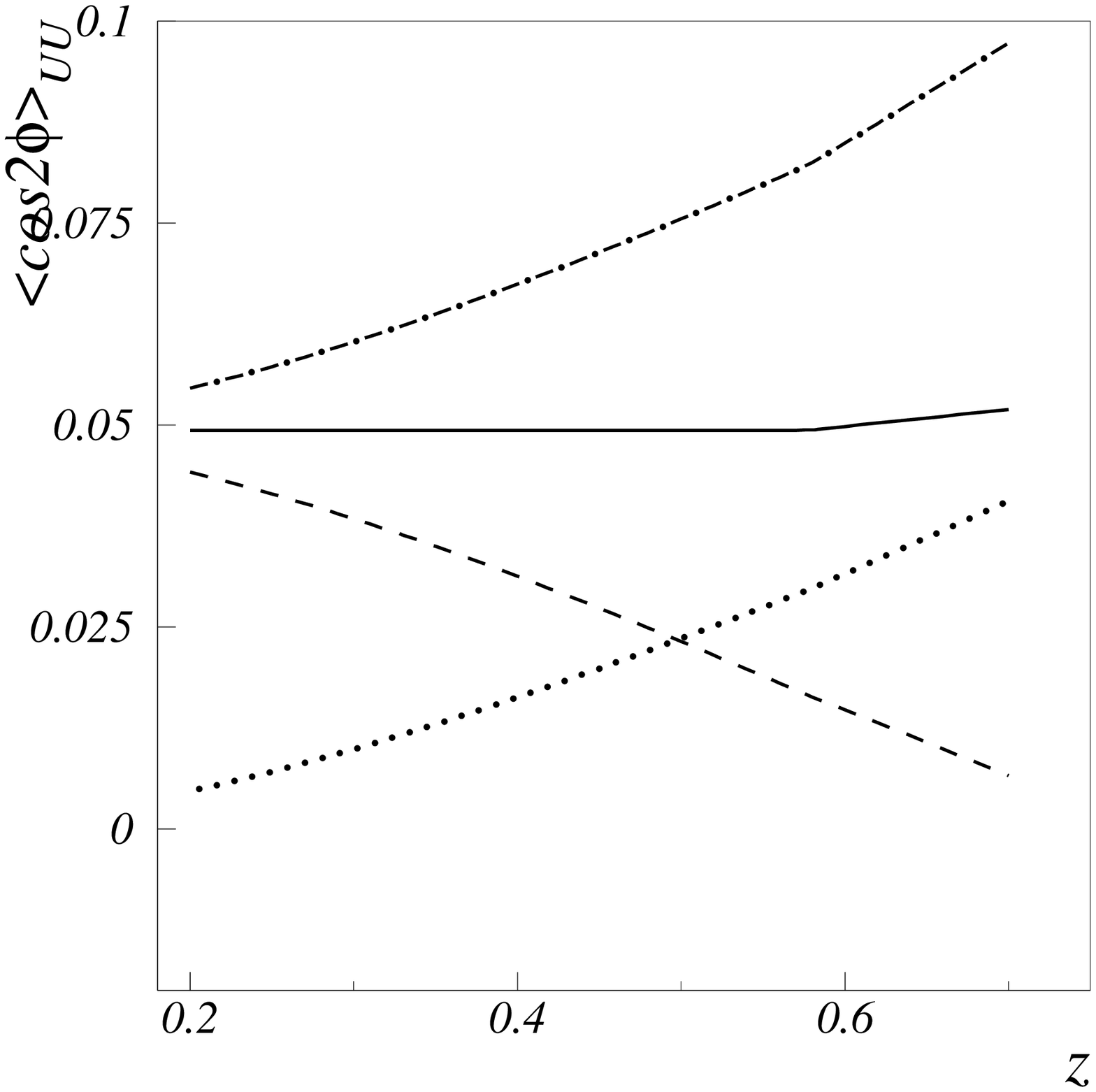}
\caption{Left panel: $\nu$ plotted as a function of $x$ for 
 $s=50\ {\rm GeV}^2$ $q_T$ ranging from 2 to 4 GeV. Right panel:
The $z$-dependence of the $\cos2\phi$ asymmetry at HERMES
kinematics. }
\label{nux}
\end{center}
\end{figure}

The angular asymmetries that 
arising in  {\em unpolarized} Drell-Yan 
scattering~\cite{cs} {\small ($\bar{p}+p \to \mu^- \mu^+ +X$)} 
and SIDIS~\cite{cahn} {\small( $e+p \to e'hX$)}
 are derived from the differential cross section expressions:
{\small
\be
\frac{1}{\sigma}\frac{d\sigma}{d\Omega}&=&
\frac{3}{4\pi}\frac{1}
{\lambda+3}\left( 1+\lambda\cos^2\theta + \mu \sin^2\theta \cos\phi 
+ \frac{\nu}{2} \sin^2\theta \cos 2\phi\right)\, ,
\nn
&&\frac{d\sigma}{dxdydzdP^2_{h\perp}d\phi_h}=  A+B+C\cos\phi+D\cos2\phi\, .
\label{cross}
\ee}
In the Drell-Yan process
the angles refer to the lepton pair orientation in their rest 
frame relative to the boost direction and the initial hadron's 
plane~\cite{cs}.  $\lambda, \mu, \nu$ depend on  $s, x, m_{\mu\mu}^2, q_T$.
In SIDIS,  the azimuthal angle  refers to the 
relative angle between the hadron production plane and the
lepton scattering  plane. $A, \ B, \ C$, 
and $D$ are functions of $x, y, z, Q^2, |\bm{P}_{h\perp}|$.    
$\nu$ is given by~\cite{boerich}
{\small
\be
\nu_2 = \frac{\sum_a e^2_a {\FF} 
\left[ 
(2 \hat{\bm{h}}\cdot \bm{k}_{\perp }\cdot\hat{\bm{h}}\cdot 
\bm{p}_{\perp }
 - \bm{p}_\perp \cdot \bm{k}_\perp)
h_1^\perp(x, \bm{k}_\perp)\bar{h}_1^\perp(\bar{x}, \bm{p}_\perp)/
(M_1 M_2)\right]
}
{\sum_a e_a^2 {\FF} 
\left[f_1(x, \bm{k}_\perp) \bar{f}_1(\bar{x}, \bm{p}_\perp)\right]}
\nnn
\ee}
where ${\FF}$ is the convolution integral. 
Collins and Soper~\cite{cs} pointed out,  well before 
$h_1^{\perp}$ was identified, there 
is a higher twist $T$-even contribution to the $\cos2\phi$ asymmetry
{\small
\bea
\nu_4=\frac{\frac{1}{Q^2}\sum_a e^2_a{\FF}
\left[2\left(\hat{\bm{h}}\cdot\left(\bm{k}_\perp -\bm{p}_\perp \right)\right)^2
-\left(\bm{k}_\perp-\bm{p}_\perp\right)^2
f_1(x, \bm{k}_\perp) \bar{f}_1(\bar{x}, \bm{p}_\perp)
\right]}
{\sum_a e_a^2 {\FF}\left[ f_1(x, \bm{k}_\perp) 
\bar{f}_1(\bar{x}, \bm{p}_\perp)\right]}.
\nnn
\eea}
\begin{figure}[h]
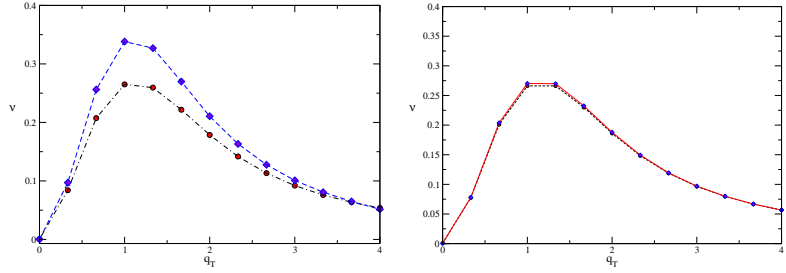

\begin{center}
\includegraphics[width=5.0cm]{nu_qt_50.eps}
\hskip 0.2cm
\includegraphics[width=5.0cm]{nu_qt_500.eps}
\caption{\label{nuqt}{\scriptsize 
$\nu$ plotted as a function of $q_T$ for 
 $s=50\ {\rm GeV}^2$ and $x$ in the range $0.2-1.0$, and  $q$ ranging from
$2.5-5.0 \ {\rm GeV}$: Right panel: 
$s=500\ {\rm GeV}^2$ and $q$ from $4.0-8.6 \ {\rm GeV}$.}}
\end{center}
\end{figure}
It is not small at center of mass energies of $50\ {\rm GeV}^2$. 
We estimate the leading twist $2$ and twist $4$ contributions~\cite{hix}. 
  In Fig.~\ref{nuqt},
at center of mass energy of $s=50\ {\rm GeV}^2$,  
the $T$-odd portion contributes about $28\% $ with an
additional $10\%$ from the
sub-leading  $T$-even piece.
The distinction between
the leading order $T$-odd and  sub-leading order 
$T$-even contributions diminish
at center of mass energy of $s=500\ {\rm GeV}^2$. 
In  Fig.~\ref{nux}, $\nu$ is plotted versus $x$ at $s=50\ {\rm GeV}^2$, 
where $q_T$ ranges from 2 to 4 GeV.  Again the higher twist contribution is
significant.
In SIDIS  a leading effect enters the $\cos 2\phi$ asymmetry with $h_1^\perp$ 
convoluted with the Collins function, $H_1^\perp$~\cite{colnpb}. 
The $\langle \cos2\phi \rangle$ from ordinary sub-leading 
$T$-even and leading $T$-odd (up to a sign) contributions 
to order $1/Q^2$ is given by
{\small
\begin{equation}
{\langle \cos2\phi \rangle}_{UU} 
=\frac{2\frac{\langle k^2_{\perp }\rangle}{Q^2} (1-y) f_1(x)D_1(z) 
\pm 8 (1-y) h^{\perp (1)}_1(x) H^{\perp (1)}_1(z)}{ \bigg [ 1+{(1-y)}^2 + 
2\frac{\langle k^2_{\perp }\rangle}{Q^2} (1-y) \bigg ] f_1(x) D_1(z)}.
\label{C2PHI}
\end{equation}}
The $z$-dependence of this asymmetry at HERMES kinematics~\cite{hermes} 
are shown in the right panel of Fig.~\ref{nux}. The 
full and dotted curves correspond to the $T$-even and $T$-odd terms in the
asymmetry, respectively. The dot-dashed and dashed curves are the sum and
the difference of those terms. One can 
see that the double 
$T$-odd asymmetry behaves like $z^2$, while the $T$-even 
asymmetry is flat in the whole range of $z$.  
Thus, aside from the
competing $T$-even effect,  the experimental 
observation of a strong $z$-dependence 
would indicate the presence of $T$-odd structures in {\it unpolarized}  
SIDIS implying that novel transversity properties of the  nucleon 
can be accessed  without invoking target polarization.     

We conclude that $T$-odd  correlations of intrinsic transverse
quark momentum and transverse spin of reacting particles  are intimately
connected with studies of the $\cos 2\phi$
azimuthal asymmetries in Drell-Yan and SIDIS.

{\scriptsize
Work done in collaboration with G. R. Goldstein and K. A. 
Oganessyan.  Acknowledgments  to F. Pijlman, A. Metz and R. Seidl 
for useful discussions and the organizers of {\em SPIN04}.  }

\end{document}